# A $4,000 Workstation for Mammalian Genome Assembly with Long Reads

Hikoyu Suzuki [1] and Norichika Ogata [1],*

[1] *Nihon BioData Corporation, 3-2-1 Sakado, Takatsu-ku, Kawasaki, Kanagawa 213-0012, Japan*

**Abstract.** Long-read sequencing has enabled the *de novo* assembly of several mammalian genomes, but with high cost in computing. Here, we demonstrated *de novo* assembly of mammalian genome using long reads in an efficient and inexpensive workstation.

## 1. Introduction

Long reads derived from single molecule real-time (SMRT) sequencing provide useful pieces of information for high-quality genome assembly. Because the sequencing error rate is quite high, long reads have been used in combination with accurate Illumina reads (hybrid assembly). Later, however, the successes of non-hybrid assembly methods were reported [1-3]. These non-hybrid assembly methods enable to obtain high-quality genome assemblies, whereas the computing costs were high. Recently, non-hybrid assembly methods were applied to characterization of mammalian genomes [4-7]. In these studies, abundant computer resources were supposed to be spent. For example, "an NFS-based computing clusters" was used [6]. In other cases, "the computational resources and staff expertise provided by the Department of Scientific Computing at the Icahn School of Medicine at Mount Sinai" [4] or "used the computational resources of the Biowulf system at the National Institutes of Health" [7]. It was also said that "Google used 405,000 CPU hours so assemble a human genome from PacBio data" [8]. Today, if we had 405,000 CPU hours on a cloud computing service, it would cost over 50,000 USD. It is clear that both inexpensive long-read sequencing technologies and assembly methods without clustered computing resources allow individual laboratories to obtain high-quality genome assemblies. Canu [9], the latest assembler for long reads, was expected to require much less computing resources. Here, we developed an inexpensive workstation and demonstrated *de novo* assembly of mammalian genome by Canu.

## 2. Materials and Methods

### 2.1. A workstation

We developed a dual-processor workstation with two CPUs, Xeon E5-2620 v4 (8 cores, 2.1 GHz, 20 MB cash, 45,996 JPY, Intel, Santa Clara, CA), a Motherboard, X10DAi (SSI-EEB, dual-LGA2011-v3, Intel C612 Chipset, 62,531 JPY, Super Micro Computer Inc., San Jose, CA), a video card, GF-GTX750TI-LE2GHD ( 640 cores, 1020 MHz, 2GB, GDDR5, 9,826 JPY, Kuroutoshikou, Japan), 8 RAMs, KVR21R15D4K4/128 (32GB, DDR4, 2133 MHz, ECC, 24,980 JPY, Kingston Technology, Fountain Valley, CA), a HDD, MD04ACA200 (3.5 inch, 2 TB, 7200 rpm, 128 MB cash, SATA 6 Gbps, 7,150 JPY, TOSHIBA, Tokyo, Japan), two HDDs, MD04ACA400 (3.5 inch, 4 TB, 7200 rpm, 128 MB cash, SATA 6 Gbps, 12,798 JPY, TOSHIBA, Tokyo, Japan), a SSD, MX300 (2.5 inch, 525 GB, SATA 6 Gbps, 16,598 JPY, Micron Technology, Inc., Boise, ID), two CPU coolers, SST-AR08 (90 mm fan, LGA2011-v3, 3,800 JPY, SilverStone Technology Co., Ltd, New Taipei, Taiwan), CPU grease, MX-4/4g (4 g, 886 JPY, ZAWARD CORPORATION, Tokyo, Japan), a power source, SST-ST75F-P (ATX 750 W PLUS SILVER, 13,000 JPY, SilverStone Technology Co., Ltd, New Taipei, Taiwan) and a case, SST-GD07B/B (SSI-EEB, 120 mm side fun, 120 mm bottom fun, 22,846 JPY, SilverStone Technology Co., Ltd, New Taipei, Taiwan) (Table 1). The total price of workstation parts was approximately 4,000 USD (457,869 JPY, 17 March 2017).

---

*Corresponding author. E-mail: norichik@nbiodata.com

*2.2. Softwares*

Operation system was Ubuntu 14.04 LTS. Canu v1.4 was installed.

*2.3. Long reads*

Approximately 78-fold long reads of mammalian genome were obtained from PacBio RS II. Median length was 6k and mean length was 7k. Shannon entropy of read length was 24.4 shannon.

*2.4. De novo assembly by Canu*

We executed Canu in our workstation with optional parameters 'minReadLength=1000 minOverlapLength=750 genomeSize=2500m maxThreads=16 -pacbio-raw' and other default parameters. We performed assembly using whole sequencing data (78-fold) and partial sequencing data (34-fold).

*2.5. Comparison of assemblies*

To evaluate the genome assembly, we compared with other two draft genomes. One was assembled by ALLPATH-LG [10] using only Illumina reads. The other was gap-closed by PBJelly [11] using whole PacBio reads based on the Illumina-based assembly. Relative values of maximum length, N10-90, total number and total length of contigs/scaffolds were calculated based on the data from hybrid-assembly using both Illumina and PacBio reads.

## 3. Results and Discussion

*3.1. Real computing time*

Canu was executed in 16 parallel threads. Assembly using whole sequence data (78-fold) was finished in 29 days. Assembly using partial sequence data (34-fold) was finished in 22 days.

*3.2. Comparison of assemblies*

Assembly using whole sequencing data (78-fold) was better than Illumina-based draft genome. Interestingly, it was also better than hybrid-assembled genome particular in max length, N10-30, mean length and total number of contigs (Figure 1). This data might reflect some restriction of Illumina-based assembly method. Partial sequencing data (34-fold) was not enough to obtain high-quality mammalian genome assembly. It was realized that the quality of assembly could be remarkably improved by only adding reads.

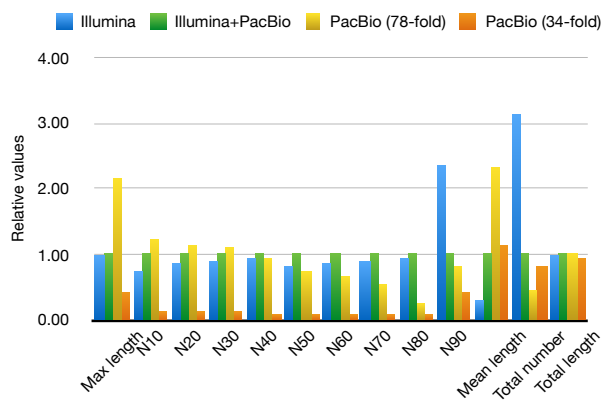

Figure 1. Comparison of Assembly

## 4. Conclusion

In this study, we could obtain high-quality mammalian genome assembly by Canu, executed in an inexpensive workstation within a month. Our demonstration shows that clustered computing systems are not necessarily required for even mammalian genome assembly, however, the computing costs should be still improved. Further reduction of computing costs for assembly is expected in the near future.

Table 1. Parts list of the dual-processor workstation

| Parts | Product Name | Spec | Price (JPY) | # |
|---|---|---|---|---|
| CPU | Xeon E5-2620 v4 | 8 cores, 2.1 GHz, 20 MB cash | 45,996 | 2 |
| motherboard | X10DAi | SSI-EEB, dual-LGA2011-3, Intel C612 Chipset | 62,531 | 1 |
| video card | GF-GTX750TI-LE2GHD | 640 cores, 1020 MHz, 2GB, GDDR5 | 9,826 | 1 |
| RAM | KVR21R15D4K4/128 | 32GB, DDR4, 2133 MHz, ECC | 24,980 | 8 |
| HDD | MD04ACA200 | 3.5 inch, 2 TB, 7200 rpm, 128 MB cash, SATA 6 Gbps | 7,150 | 1 |
| HDD | MD04ACA400 | 3.5 inch, 4 TB, 7200 rpm, 128 MB cash, SATA 6 Gbps | 12,798 | 2 |
| SSD | MX300 | 2.5 inch, 525 GB, SATA 6 Gbps | 16,598 | 1 |
| CPU cooler | SST-AR08 | 90 mm fan, LGA2011-v3 | 3,800 | 1 |
| CPU grease | MX-4/4g | 4 g | 886 | 1 |
| Power source | SST-ST75F-P | ATX 750 W PLUS SILVER | 13,000 | 1 |
| Case | SST-GD07B/B | SSI-EEB, 120 mm side fun, 120 mm bottom fun | 22,846 | 1 |